\documentstyle[12pt,aasms]{article}
\input epsf
\tighten
\received{}
\accepted{}
\journalid{}{}
\articleid{}{}
\slugcomment{}
\lefthead{Macri {\it et al.}}
\righthead{Cepheid distance to \n4535}
\begin{document}
\def \Vmod    {31.30\pm0.05}
\def \Imod    {31.19\pm0.04}
\def \muo     {31.03\pm0.05{\rm\ (random)} \pm0.26{\rm\ (systematic)}}
\def \dist    {16.1\pm0.4{\rm\ (random)} \pm1.9{\rm\ (systematic)}}
\def \EVI     {0.11\pm0.02}
\def \VmodDo  {31.25\pm0.04}
\def \ImodDo  {31.15\pm0.04}
\def \muoDo   {31.01\pm0.05{\rm\ (random)} \pm0.26{\rm\ (systematic)}}
\def \distDo  {15.9\pm0.4{\rm\ (random)} \pm1.9{\rm\ (systematic)}}
\def \EVIDo   {0.10\pm0.02}
\def \nc{50 }
\def \ni{39 }
\def \ran{14--72 }
\def \vi{{\it V--I\/}}
\def \bv{{\it B--V\/}}
\def \n4535{NGC$\,$4535}
\def \Deg{${}^\circ$\llap{.}}
\def \Sec{${}^{\prime\prime}$\llap{.}}

\let\footnotesize=\small
\vspace*{-1.9truein}
\title{The Extragalactic Distance Scale Key Project XVIII. \\
The Discovery of Cepheids and a New Distance to \n4535 \\
Using the Hubble Space Telescope$^\dagger$}

\author{
L.M.~Macri\altaffilmark{1},
J.P.~Huchra\altaffilmark{1},
P.B.~Stetson\altaffilmark{2},
N.A.~Silbermann\altaffilmark{3},
W.L.~Freedman\altaffilmark{4},
R.C.~Kennicutt\altaffilmark{5},
J.R.~Mould\altaffilmark{6},
B.F.~Madore\altaffilmark{3},
F.~Bresolin\altaffilmark{7},
L.~Ferrarese\altaffilmark{8},
H.C.~Ford\altaffilmark{9},
J.A.~Graham\altaffilmark{10},
B.K.~Gibson\altaffilmark{11},
M.~Han\altaffilmark{12},
P.~Harding\altaffilmark{5},
R.J.~Hill\altaffilmark{13},
J.G.~Hoessel\altaffilmark{14},
S.M.G.~Hughes\altaffilmark{15},
D.D.~Kelson\altaffilmark{10},
G.D.~Illingworth\altaffilmark{16},
R.L.~Phelps\altaffilmark{3},
C.F.~Prosser\altaffilmark{17}$^*$,
D.M.~Rawson\altaffilmark{6}$^*$,
A.~Saha\altaffilmark{18}$^\ddagger$,
S.~Sakai\altaffilmark{17}
\& A.~Turner\altaffilmark{5}}

\altaffiltext{1}{Harvard-Smithsonian Center for Astrophysics, Cambridge,
MA 02138, USA}

\altaffiltext{2}{Dominion Astrophysical Observatory, Victoria, BC V8X
4M6, Canada}

\altaffiltext{3}{IPAC, California Institute of Technology, Pasadena, CA
91125, USA}

\altaffiltext{4}{Carnegie Observatories, Pasadena, CA 91101, USA}

\altaffiltext{5}{Steward Observatory, University of Arizona, Tucson, AZ
85721, USA}

\altaffiltext{6}{MSSSO, Australian National University, Weston Creek,
ACT 2611, Australia}

\altaffiltext{7}{European Southern Observatory, Garching b. M\"unchen,
D-85478, Germany}

\altaffiltext{8}{Department of Astronomy, California Institute of
Technology, Pasadena, CA 91125, USA}

\altaffiltext{9}{John Hopkins University and Space Telescope Institute,
Baltimore, MD 21218, USA}

\altaffiltext{10}{DTM, Carnegie Institution of Washington, Washington,
DC 20015, USA}

\altaffiltext{11}{CASA, University of Colorado, Boulder, CO 80309, USA}

\altaffiltext{12}{Avanti Corporation, Fremont, CA 94538, USA}

\altaffiltext{13}{University of Wisconsin, Madison, WI, 53706, USA}

\altaffiltext{14}{Lab. for Astronomy and Solar Physics, NASA GSFC,
Greenbelt, MD 20771, USA}

\altaffiltext{15}{Institute of Astronomy, Unversity of Cambridge,
Cambridge CB3 0HA, UK}

\altaffiltext{16}{Lick Observatory, University of California, Santa
Cruz, CA 95064, USA}

\altaffiltext{17}{National Optical Astronomy Observatories, Tucson, AZ
85726, USA}

\altaffiltext{18}{Space Telescope Science Institute, Baltimore, MD
21218, USA}

\renewcommand{\thefootnote}{\fnsymbol{footnote}}
\footnotetext[2]{Based on observations with the NASA/ESA
{\it Hubble Space Telescope}, obtained at the Space Telescope Science
Institute, operated by AURA, Inc. under NASA contract No. NAS5-26555.}
\footnotetext[1]{Deceased}
\footnotetext[3]{Current address: National Optical Astronomy Observatories,
Tucson, AZ 85726, USA}

\begin{abstract}
We report on the discovery of Cepheids in the Virgo spiral galaxy
NGC$\,$4535, based on observations made with the Wide Field and Planetary
Camera 2 on board the {\it Hubble Space Telescope}. \n4535 is one of 18
galaxies observed as a part of {\it The HST Key Project on the Extragalactic
Distance Scale}, which aims to measure the Hubble constant to 10\%
accuracy. \n4535 was observed over 13 epochs using the F555W filter, and over
9 epochs using the F814W filter. The HST F555W and F814W data were
transformed to the Johnson $V$ and Kron-Cousins $I$ magnitude systems,
respectively. Photometry was performed using two independent programs, DoPHOT
and DAOPHOT~II/ALLFRAME.
\vskip 12pt 
Period-luminosity relations in the $V$ and $I$ bands were constructed
using \ni high-quality Cepheids present in our set of \nc variable
candidates.  We obtain a distance modulus of $31.02\pm0.26$~mag,
corresponding to a distance of $16.0\pm1.9$~Mpc. Our distance estimate
is based on values of $\mu_0=18.50 \pm 0.10$~mag and E(\vi) = 0.13~mag
for the distance modulus and reddening of the LMC, respectively.
\end{abstract}

\keywords{Cepheids --- distance scale --- galaxies: individual (\n4535)}

\newpage

\section{Introduction}

The Hubble Space Telescope Key Project on the Extragalactic Distance Scale
(\cite{ken95}), aims to measure the Hubble constant to an accuracy of
10\%. This is to be achieved by obtaining distances based on Cepheid variable
stars to eighteen different galaxies, which will provide a firm basis for the
calibration of several secondary distance indicators: the Tully-Fisher
relation, surface brightness fluctuations, the planetary nebula luminosity
function, the globular cluster luminosity function, type II supernova expanding
photospheres, and type Ia supernova peak magnitudes (see \cite{jac92} for a
review of these methods). 

This paper presents the discovery of \nc Cepheids and a distance determination
to \n4535. \n4535 is located at $\alpha = $ 12$^{\rm h}$ 34$^{\rm m}$ 20$^ {\rm
s}$, $\delta =$ +08$^{\circ}$ 11${\arcmin}$ 54${\arcsec}$ (J2000.0), about
1\Deg8 from NGC$\,$4472, the brightest galaxy in the Virgo cluster and the
center of the Virgo S cloud.  Its morphological classification is SAB(s)c
I/I-II (\cite{RC3}) or SB(s)c I.3 (\cite{RSA}) and its inclination angle is
47$^{\circ}$ (\cite{RC3}), making it a suitable calibrator of the Tully-Fisher
relation. Previous estimates of the distance modulus to this galaxy include
$32.27\pm0.16$~mag (based on angular diameters of high-luminosity field spiral
galaxies, \cite{san93}), $30.75\pm0.2$~mag (based on multi-band Tully-Fisher
studies of the Virgo and Ursa Major clusters, \cite{pie88}) and $30.46\pm0.5$
~mag (based on corrected isophotal diameters, total magnitudes, and corrected
luminosity indexes, \cite{vau79}).

We describe the observations and preliminary reductions in \S 2.  The
photometry and calibration of the instrumental magnitudes are discussed in \S
3. The search for Cepheids and the resulting set of variables are described
in \S 4. Period-luminosity relations and distance moduli are presented and
discussed in \S 5. A discussion of the distance to the Virgo Cluster is
presented in \S 6, and our conclusions are given in \S 7.

\section{Observations and Reductions}

\subsection{Observations}

\n4535 was observed by the Wide Field and Planetary Camera 2 (WFPC2) (\cite
{bur94}) on board HST on twelve epochs between 1996 May 24 and August 7.  The
observations were performed using the F555W (approximately Johnson $V$) and the
F814W (approximately Kron-Cousins $I$) filters. Eight of the epochs contained
one hour of F555W data (3$\times$1200$\,$s images) and 3900$\,$s of F814W data
(3$\times$1300s images) each, while the other four consisted of 5000$\,$s of
F555W data (2$\times$1200$\,$s + 2$\times$1300$\,$s images) each. A dithering
pattern was applied between epochs in order to increase the sampling of the
PSF. A pre-visit observation was performed on 1994 June 9, returning one hour
of data in each filter.  The aim of this early observation was to provide a
long-baseline epoch to reduce aliasing effects for the longest-period Cepheids.
All observations were made with the telescope guiding in fine lock with a
stability of approximately 3 mas. The gain and readout noise were 7 e$^{-}$/DN
and 7 e$^{-}$, respectively. The CCD was operated at a temperature of
--88$^\circ\,$C for all observations. A log of the observations is presented in
Table~1. A V-band image of \n4535 (obtained at the Whipple Observatory 1.2-m
telescope) is shown in Figure 1 with the WFPC2 mosaic overlaid at the correct
scale and orientation. This mosaic is shown in greater detail in Figure~2.

This data set is slightly different from the standard set obtained for other
Key Project galaxies. The 1994 data were obtained at a roll angle of
293$^\circ$, while the 1996 observations were performed at an angle of
288$^\circ$. This group of images shows a steady rise of the background level
(from 95 DN to 210 DN in the PC chip), due to a continued decrease of the solar
angle from 120$^\circ$ to 50$^\circ$. Lastly, a safing event prevented the
next-to-last observation from being performed at its scheduled time and
therefore slightly disrupted the intended sampling. Generally, these
differences do not impact our ability to find and measure Cepheids.

The sampling strategy, as discussed by Freedman et al.~(1994), has been
designed specifically for the Key Project with the purpose of maximising the
probability of detecting Cepheids with periods ranging from 10 to 75 days. It
follows a power-law distribution in time which provides an optimum sampling of
the light curves of Cepheids in this period range and reduces the risk of
aliased detections. Figure~3 shows the sampling variance for the $V$-band
observations. Lower values of variance (towards the top of the plot) indicate
more uniform phase coverage and higher probabilities of detection, while higher
values of variance indicate less uniform phase coverage and correspondingly 
lower probabilities of detection.

\subsection{Data Reductions}

The HST data have been calibrated using the pipeline processing at the Space
Telescope Institute (STScI). The full reduction procedure (\cite{hol95a})
consists of: a correction of A/D conversion errors, the subtraction of a bias
level for each chip, the subtraction of a superbias frame, the subtraction of a
dark frame, a correction for shutter shading, and a division by a flat
field. The names of the STScI reference files used for this calibration are
listed in the notes to Table 1. Furthermore, each of the frames was corrected
for vignetting and geometrical distortions in the WFPC2 optics (the latter was
done using files kindly provided by J.~Holtzman). Lastly, the images were
multiplied by four and converted to two-byte integers, to reduce disk usage and
allow image compression. This conversion led to an effective readout noise of 4
DN and a gain of 1.75e$^-$/DN.

No correction for charge transfer inefficiency was applied since CTI affects
images with backgrounds lower than 30e$^-$ (Holtzman et al.~1995b) and our data
always exceeded this value. Similarly, the observations of Leo I, NGC$\,$2419
and Pal 4 used at various stages of the DoPHOT and ALLFRAME reductions have
high background levels and are not affected by CTI. Following Hill et
al.~(1998), we use the zeropoints of the WFPC2 which are appropriate for
observations with long exposure times and high background levels. A more
complete description of the calibration steps can be found in Hill et
al.~(1998) and Stetson et al.~(1998).

\section{Photometry and Calibration}

\subsection{Photometric Reductions}

Photometry of \n4535 was performed using two independent programs: a version of
DoPHOT (\cite{sch93}) modified to deal with HST WFPC2 data
(\cite{sah96a}), and DAOPHOT~II/ALLFRAME (\cite{ste94}). The reader is referred
to those respective publications for a detailed description of each program,
and to Stetson et al.~(1998) for comprehensive overviews of the DoPHOT and
ALLFRAME reduction strategies.  The motivation for the use of two independent
programs is that each one performs photometry using independent techniques, and
a comparison of results yields a fundamental check on the validity of the
software and the photometry. This provides a powerful tool for detecting the
presence of errors and biases which might go undetected if only one program
were used.

\subsection{Calibration}

The calibration of photometry for \n4535 involves the transformation of $F555W$
and $F814W$ magnitudes into the standard system (Johnson $V$ and Kron-Cousins
$I$). This is done via the following equations, based on Equation (8) of
Holtzman et al. (1995b):

\vskip -18pt
\begin{eqnarray}
V&=&F555W+T_{1,F555W} (V-I) + T_{2,F555W} (V-I)^2+Z_{F555W}+2.5\log GR_i\\
I&=&F814W+T_{1,F814W} (V-I) + T_{2,F814W} (V-I)^2+Z_{F814W}+2.5\log GR_i
\end{eqnarray}

\noindent{where $F555W$ and $F814W$ are $-2.5\log({\rm DN\ s}^{-1})$, the
various $T$ coefficients are the color term transformations, the $Z$ terms
are the zeropoint coefficients, and the $GR_i$ terms correct for changes in
the gain state of the camera from 14~e$^-$/DN for the calibration data to
7~e$^-$/DN for the \n4535 observations. The gain ratio terms (from
\cite{hol95b}) are 1.987, 2.003, 2.006 and 1.955 for the PC1, WF2, WF3 and
WF4 chips, respectively, and have uncertainties of order 1\%.}

The DoPHOT and ALLFRAME transformations from the instrumental to the standard
system are slightly different, because the DoPHOT zeropoints are based on the
``ground system'' while the ALLFRAME zeropoints are based on the ``flight
system.'' Therefore, the $T_1$ terms used in the DoPHOT transformation are
those from Table 3 of Holtzman et al. (1995b) ($T_{1,F555W} = -0.045\pm0.003$,
$T_{1,F814W} = -0.067\pm0.003$) whereas the $T_1$ terms used in the ALLFRAME
transformation are those from Table 7 of the same paper ($T_{1,F555W} =
-0.052\pm0.007$,$T_{1,F814W} = -0.062\pm0.009$). The $T_2$ terms are
identical in both systems ($T_{2,F555W} = 0.027\pm0.002$, $T_{2,F814W} =
0.025\pm0.002$).

\subsection{Comparison of DoPHOT and ALLFRAME Photometry}

A comparison of the DoPHOT and ALLFRAME photometric systems was conducted for
all chip/filter combinations using a visually-selected subset of stars having
no bright neighbours within five pixels. Their coordinates (obtained using
IRAF's {\tt stsdas.hst\_calib.wfpc.metric}) and $V$ and $I$ DoPHOT and ALLFRAME
magnitudes are listed in Tables A3-A6 for future comparisons of our photometry
with other observations of \n4535. Figure~4 shows plots of $\Delta{\rm mag}$
(DoPHOT-ALLFRAME) versus DoPHOT magnitude for these stars. The mean offsets
between the two photometric systems for each chip and filter combination are
listed in Table A2. Based on our results, we estimate a photometric uncertainty
of $\pm0.04$~mag in both bands for bright ($m < 24$~mag), isolated stars.

\section{Identification of Cepheid Candidates}

\subsection{Search for Cepheids}

The search for Cepheid variables using the DoPHOT and ALLFRAME data was carried
out in three steps. The first step was a classification of all objects as
either variable or non-variable.  The second step consisted of a periodicity
test of the variables and a preliminary determination of periods.  These two
steps are described in detail in Stetson et al.~1998 and will not be repeated
here.  The third step involved the use of template-fitting procedures
(\cite{ste96}) to identify those periodic variables that had Cepheid-like light
curves, refine their periods and calculate mean magnitudes.

In general, observations of Cepheids with HST suffer from sparse sampling and
coverage of the phased lightcurve. Therefore, the adopted approach for the
determination of mean magnitudes is the use of template-fitting procedures. All
the $V$ and $I$ data points of each variable were fit to a light curve
template, described in detail by Stetson (1996). The $V$-band template is a
five-term Fourier expansion, while the $I$-band uses four terms.  The templates
were derived from a sample of 114 Milky Way and Magellanic Cepheids with
periods between 7 and 100 days. The template-fitting procedure refined the
period estimates and calculated their uncertainties. Furthermore, it performed
a numerical integration of the fitted light curves to calculate
mean $V$ and $I$ magnitudes and their uncertainties.

\subsection{The Cepheids Found in \n4535}

The DoPHOT and ALLFRAME variable searches produced a total of \nc Cepheid
candidates (identified as C01-C\nc in order of descending period). Table~2
lists, for each candidate, its identification number, the chip in which it is
found, its J2000.0 coordinates (obtained with IRAF's {\tt
stsdas.hst\_calib.wfpc.metric}) and a quality grade. The quality grade was
independently determined by five of us, who graded each candidate on a scale of
1 (perfect) to 4 (very poor) on three aspects: the DoPHOT light curves, the
ALLFRAME light curves, and the degree of crowding in the vicinity of each
star. Only those stars with a quality grade less than or equal to 2.0 were used
in the final fits, and they are marked with an asterisk next to their IDs;
there are \ni such candidates in our set. Table~3 lists, for each candidate,
the DoPHOT and ALLFRAME periods and mean magnitudes for the $V$ and $I$ bands.

A comparison of the DoPHOT and ALLFRAME mean $V$ and $I$ Cepheid magnitudes for
variables in each of the four chips is presented in Figure~5. The mean
offsets and {\it rms} dispersions are listed in Table A2. While the small
number of variables in the PC1 and WF2 may not allow a good determination of
the offsets in those chips, the Cepheids in the WF3 and WF4 chips indicate an
agreement in the mean of about 0.05~mag between the photometry
programs. However, the distributions become asymmetric as we go towards fainter
magnitudes (clearly seen in the case of of the WF3 chip) and magnitude
differences grow to 0.15-0.3 mag. This phenomenon has been seen
in many of the galaxies studied in this project, and artificial star tests
(\cite{fer99}) are being conducted to try to understand its origin.

The Cepheids are identified in the field of each of the WFPC2 chips in
Figures~6a-d. Finding charts for each of the stars are displayed in
Figures~7a-b; the charts encompass a 50 by 50 pixel area around each variable
(i.e., 5\arcsec\ x 5\arcsec\ for the WF and 2\Sec15$\times$2\Sec15 for the
PC). $F555W$ and $F814W$ light curves based on the DoPHOT data are presented in
Figures~8a-i.

The individual photometric measurements are listed in Tables~A7 and A8 for
DoPHOT F555W- and F814W-band data, and in Tables~A9 and A10 for ALLFRAME $V$-
and $I$-band data, respectively (for those epochs without F814W data, the
ALLFRAME transformation to the standard system was carried out by assuming a
\vi = 0 color).

\section{Period-Luminosity Relations and Distance Moduli}
\subsection{Methodology}
The method used to derive $V$- and $I$-band apparent distance moduli is the
same one that was used in other papers in this series. The period-luminosity
relations of LMC Cepheids (Madore \& Freedman 1991) are scaled to an assumed
true distance modulus of $\mu_{0,{\rm LMC}}$ = 18.50$\pm$0.10~mag (total
uncertainty) and corrected for an estimated average line-of-sight reddening of
E(\vi)$_{\rm LMC}=0.13$~mag (\cite{bes91}).

The period-luminosity relations are:
\vskip -24pt
\begin{eqnarray}
M_V&=&-2.76(\pm0.11)\left[\log{\rm P}\left({\rm d}\right)-1.0\right]
      -4.16(\pm0.05), \\
M_I&=&-3.06(\pm0.07)\left[\log{\rm P}\left({\rm d}\right)-1.0\right]
      -4.87(\pm0.03).
\end{eqnarray}

In fitting the data from \n4535, we follow the procedure given in Freedman et
al.~(1994). We fix the slope to that given in the above equations and obtain
the zeropoint by minimizing the unweighted {\it rms} dispersion. By using the
slopes given above, we minimize any bias that would arise from incompleteness
at faint magnitudes in the \n4535 data set.

The magnitude shifts resulting from this method are converted to $V$- and
$I$-band apparent distance moduli by substracting the magnitude zeropoints of
Equations (5) and (6) and extrapolating to $\log P=0$.  A reddening correction
due to the presence of interstellar dust must be applied in order to obtain
the true distance modulus (see \cite{mad91} for a complete description of this
correction). In determining the reddening-corrected true modulus, a
reddening-corrected magnitude $W$ = $V$ -- 2.45(\vi) is calculated for each
Cepheid and differenced against the absolutely calibrated version of the
$W-\log P$ relation (see \cite{mad82} for details).

\subsection{Results}
A subset of high-quality Cepheids (quality grade less than or equal to 2.0)
were selected for P-L fitting; there are \ni such stars in the data set. P-L
relations in the $V$ and $I$ bands are plotted in Figures~9 and 10 for DoPHOT
and ALLFRAME data, respectively. Filled circles are used to plot the Cepheids
in the selected subset, while the remainder are represented by open
circles. Figure~11 shows the location of all Cepheids in a color-magnitude
diagram of \n4535 stars, using the same plotting scheme.

The fits to the P-L relations resulted in $V$- and $I$-band apparent distance
moduli of $\VmodDo$~mag and $\ImodDo$~mag for the DoPHOT data and $\Vmod$~mag
and $\Imod$~mag for the ALLFRAME data (all of the uncertainties quoted here are
fitting errors only). Assuming a mean LMC reddening of E(\vi) = 0.13~mag, we
obtain mean reddening values of E(\vi) = $\EVIDo$~mag and $\EVI$~mag for the
DoPHOT and ALLFRAME data, respectively.

The apparent distance moduli were combined to obtain true distance moduli of
$\muoDo$~mag and $\muo$~mag, corresponding to distances of $\distDo$~Mpc and
$\dist$~Mpc, for the DoPHOT and ALLFRAME data, respectively. If the complete
set of \nc Cepheid candidates is used, the DoPHOT distance modulus changes by
$-0.03$~mag and the ALLFRAME distance modulus changes by $-0.01$~mag.

\subsection{Error Budget}
The error bugdet for our distance determination is summarized in Table 4, and
explained in greater detail here. The first source of uncertainty is related to
the LMC Cepheid P-L relations. The largest contribution to this uncertainty
comes from the uncertainty in the distance modulus of the LMC; recent estimates
range from $18.22\pm0.13$~mag (\cite{uda98}) to $18.70\pm0.10$~mag
(\cite{fea97}).  In keeping with other papers in this series, we adopt
$\pm0.10$~mag as a likely value for the $1-\sigma$ uncertainty in the distance
modulus. The other two (minor) contributions to our first uncertainty come from
the zeropoint uncertainties of the Madore \& Freedman (1991) $V$- ($\pm
0.05$~mag) and $I$-band ($\pm 0.03$~mag) P-L relations. These two sources of
uncertainty will be reduced even further in the near future, once the improved
LMC P-L relations of Sebo et al. (1999) are published, but the largest
uncertainty, that of the LMC distance, will probably be with us for some time
to come. Combining these three sources of error in quadrature, we arrive at the
total uncertainty for the absolute calibration of the Cepheid P-L relation of
$\pm 0.12$~mag. We characterize this uncertainty as ``systematic,'' since it
applies to all galaxies studied in this project. Any change in the value or the
uncertainty for the LMC distance or the P-L zeropoints will affect the derived
value and uncertainty for the Hubble constant directly.

Our second source of uncertainity is related to the absolute photometric
calibration of WFPC2. We have estimated the total uncertaintity in
the transformations from the instrumental to the standard system to be
$\pm0.05$~mag in each filter; this includes the uncertainties in the
characterization of the zeropoints, the color transformations, the camera gain
ratios, and the effects of CTE. Since our reddening-free distance modulus is
calculated using the difference between the $V$ and $I$ apparent distance
moduli, the errors must be added in quadrature and multiplied by the reddening
factor, resulting in a total error of $\pm 0.15$~mag. We characterize this
uncertainty as ``systematic,'' since any changes in WFPC2 zeropoints will
translate directly into changes in all our distance moduli and the derived
value of the Hubble constant. The recent work of Stetson (1999) will help lower
the uncertainties in the zeropoints and thus reduce this particular source of
error.

Our third source of uncertainty is related to the P-L fits of the NGC$\,$4535
set of variables. Here we have taken the {\it rms} dispersions of our P-L
relations and divided them by the square root of $N-1$, where $N=39$ is the
number of variables used in the fit. The two errors ($\pm0.05$~mag
and $\pm0.04$~mag, respectively) are partially correlated due to the
intrinsic width of the P-L relation, and therefore the total uncertainty is
$\pm0.05$~mag. We label this uncertainty as ``random,'' because it could be
further reduced if we had found more Cepheids in the galaxy.

The last and perhaps most important source of uncertainty is related to the
effect of metallicity on the Cepheid P-L relations. The LMC and NGC$\,$4535
have quite different metallicities, [O/H] $\simeq -0.4$ (\cite{ken98}) for the
former and [0/H] $\simeq 0.3$ (\cite{ken99}) for the latter.  The recent
estimate (\cite{ken98}) of the metallicity dependence of Cepheid absolute
magnitudes, $\gamma_{VI} = -0.24\pm0.16$~mag/dex, implies a correction to the
distance modulus of $+0.17\pm0.11$~mag. However, other determinations of the
metallicity effect yield coefficients closer to $-0.4$~mag/dex (Sasselov et
al.~1997, Kochanek 1997, Storm, Carney \& Fry 1999) in which case the size of
the correction would be $\simeq 0.3$~mag. This is also a ``systematic''
uncertainty because such a metallicity effect will change (by different
amounts, based on individual metallicities) the distances to all our
galaxies. We do not apply the correction to our distance modulus to be
consistent with previous papers in this series. Therefore, we combine the
correction factor of $+0.17$~mag in quadrature with the other systematic errors
to arrive at a total ``systematic'' error of $\pm0.26$~mag.  This is combined
in quadrature with the ``random'' error of $\pm0.05$~mag to arrive at a total
uncertainty for the distance modulus of $\pm0.26$~mag.

\section{Discussion}

There are now Cepheid-based distances to seven galaxies located within ten
degrees of the luminosity-weighted center of the Virgo Cluster. Table 5 lists
those distances, for which the straight average yields a mean cluster distance
of $17.3\pm3.4$~Mpc.  If we exclude the galaxy with the lowest distance
(NGC$\,$4571, \cite{pie94}) and the one with the highest distance (NGC$\,$4639,
\cite{sah97}), the average distance becomes $16.2\pm0.2$~Mpc.  This result
could indicate a nearby and compact Virgo Cluster core or it could point
towards a selection bias which tends to favor well-resolved galaxies (i.e.,
those located on the near side of the cluster) for Cepheid searches.

There is additional evidence that supports the case for the short distance to
the cluster core. B\" ohringer et al.~(1997) analyzed the ROSAT isophotes of
the cluster as well as 21-cm maps of twenty-three cluster spirals (produced by
\cite{cay90}) and found evidence of ram-pressure stripping in NGC$\,$4501 (for
which no Cepheid distance is available) and NGC$\,$4548 (for which \cite{gra98}
recently derived a distance of $16\pm2$~Mpc). B\" ohringer et al. combined the
x-ray and 21-cm data to derive a line-of-sight distance difference between M87
and NGC$\,$4548 of 0--2 Mpc.  This would place M87 at a distance of 14--16 Mpc,
since NGC$\,$4548 is falling in from behind the cluster as evidenced by its low
recession velocity and the shape of its 21-cm isophotes (\cite{cay90}). Such a
distance for M87 would be in good agreement with Jacoby et al.~(1990), who
placed the Virgo Cluster core at $14.7\pm1$~Mpc and with the results of Harris
et al.~(1998) who give a Virgo distance of $15.7\pm1.5$~Mpc. A Cepheid distance
to NGC$\,$4501 would be very helpful to tighten the constraint on the M87
distance.

\section{Conclusions}
We have discovered a total of \nc Cepheid variable candidates in the SAB(s)c
spiral galaxy NGC 4535, which is a member of the Virgo Cluster. We have fitted
standard $V$- and $I$-band period-luminosity relations to a subset of the
candidates and have obtained very similar results from our ALLFRAME and DoPHOT
analysis: a distance modulus of $31.02\pm0.26$~mag, corresponding to a distance
of $16.0\pm1.9$~Mpc, and a reddening of E(\vi)=$0.11\pm0.02$~mag.  Our distance
measurement is in good agreement with other distance determinations of spirals
in the Virgo Cluster.

\section{Acknowledgements}
We would like to thank Doug Van Orsow, the program coordinator for this Key
Project, as well as the rest of the STScI and NASA support staff that have
made this project possible. L.M.M. acknowledges support through Gemini
Fellowship No. GF-1003-95, and would like to thank K.Z.~Stanek and
M. Krockenberger for providing the template-fitting procedures used in the
DoPHOT reduction.  S.M.G.H. and P.B.S. are grateful to NATO for travel
support via a Collaborative Research Grant (960178).

The HST Key Project on the Extragalactic Distance Scale is supported by NASA
through grant GO-2227-87A from STScI. This work has made use of the NASA/IPAC
Extragalactic Database (NED), which is operated by the Jet Propulsion
Laboratory, Caltech, under contract with the National Aeronautics and Space
Administration.

\vfill\pagebreak\newpage
\begin{center}
{\bf Figure Captions}
\end{center}
\par
\noindent{Figure 1: -- A V-band wide-field image of \n4535, obtained at the
Whipple Observatory 1.2-m telescope, with the WFPC2 mosaic overlaid at its
scale and orientation (see Figure~2)}\baselineskip=18pt

\noindent{Figure 2.-- A mosaic of WFPC2 images of \n4535. Images of each chip
showing the location of the variables can be found in Figures~6a-d.}

\noindent{Figure 3.-- Sampling variance for the $V$-band observations. The
higher variance points (bottom of the plot) indicate periods at which the phase
coverage is less uniform, and a corresponding decrease in the probability of
detection. Due to the power-law distribution in time of the observations, an
enhanced performance in the detection of variables is obtained.}

\noindent{Figure 4.-- A comparison of the DoPHOT and ALLFRAME $V$ and $I$ mean
magnitudes for bright and fairly isolated stars present in each chip. Filled
and open circles represent $V$ and $I$~magnitudes, respectively. The mean
offsets and {\it rms} deviations are listed in Table A2.}

\noindent{Figure 5.-- A comparison of the DoPHOT and ALLFRAME $V$ and $I$ mean
magnitudes for Cepheids in each chip. Filled and open
circles represent $V$ and $I$ magnitudes. The mean offsets and {\it rms}
deviations are listed in Table A2.}

\noindent{Figures 6a-d.-- $V$-band images of the four WFPC2 chips. The circles 
indicate the position of each of the Cepheids, labelled as in Table~2. Each
of the images is oriented such that the bottom left corner has the pixel 
coordinates (0,0) in each image.}

\noindent{Figures 7a-b.-- Finding charts for all the Cepheids. The position of
each Cepheid is shown by the circle. The charts are 50 pixels on a side.}

\noindent{Figures 8a-i.-- Phased light curves for all the Cepheids, using
DoPHOT data. Filled and open circles represent F555W and F814W magnitudes,
respectively. Two complete cycles are plotted to facilitate ease of
interpretation.}

\noindent{Figure 9.-- Period-luminosity relations in the V (top) and I (bottom)
bands based on DoPHOT photometry. Filled circles are used to plot the subset of
high-quality Cepheids, while open circles represent the ones with bad
grades. The solid lines are the best fits and the dotted lines correspond to
the 2-$\sigma$ dispersion of the LMC period-luminosity relation of Madore
\&~Freedman (1991). Apparent distance moduli obtained from the lines of best
fit are $\VmodDo$ and $\ImodDo$ in the V- and I-bands respectively. Including a
correction due to reddening this corresponds to a distance modulus of
 $\muoDo$~mag.}

\noindent{Figure 10.-- Period-luminosity relations in the V (top) and I
(bottom) bands based on ALLFRAME photometry. Filled circles are used to plot
the subset of high-quality Cepheids, while open circles represent the ones with
bad grades. The solid lines are the best fits and the dotted lines correspond
to the 2-$\sigma$ dispersion of the LMC period-luminosity relation of Madore
\&~Freedman (1991). Apparent distance moduli obtained from the lines of best
fit are $\Vmod$ and $\Imod$ in the V- and I-bands respectively. Including a
correction due to reddening this corresponds to a distance modulus of
 $\muo$~mag.}

\noindent{Figure 11.-- Color-magnitude diagram of stars in \n4535. Filled and
open circles are used to plot the high- and low-quality Cepheids, respectively.
}

\vfill\pagebreak\newpage
\centerline{\bf Table Captions}

\noindent{Table 1: -- Log of HST WFPC2 observations of \n4535. The
Julian Date quoted is the heliocentric Julian Date at mid-exposure.}

\noindent{Table 2: -- The Cepheid candidates found in \n4535. The columns 
provide the following information: (1) the identification; (2) the chip in 
which each variable is located; (3) and (4) the right ascension and 
declination in (J2000.0) and (5) the quality index.}

\noindent{Table 3: -- The Cepheid candidates found in \n4535. The columns
provide the following information: (1) the identification; (2) the ALLFRAME
period; (3) and (4) the ALLFRAME $V$- and $I$-band mean 
magnitudes; (5) the DoPHOT period; (6) and (7) the DoPHOT $V$- and $I$-band
mean magnitudes.}

\noindent{Table 4: -- The contributions to the error budget of the 
distance modulus of \n4535.} 

\noindent{Table 5: -- Cepheid Distances to Virgo Cluster Galaxies.}

\vskip 0.5in

\centerline{\bf Appendix Table Captions}

\noindent{Table A1: -- The aperture correction and WFPC2 zero-point
coefficients used for ALLFRAME photometry.}

\noindent{Table A2: -- Mean offsets between DoPHOT and ALLFRAME magnitudes
for secondary standards and variables.}

\noindent{Tables A3-A6: -- Secondary standard photometry for Chips 1-4.}

\noindent{Table A7: -- $F555W$-band DoPHOT data for each of the Cepheids over 
the thirteen epochs.}

\noindent{Table A8: -- $F814W$-band DoPHOT data for each of the Cepheids over 
the nine epochs.}

\noindent{Table A9: -- $V$-band ALLFRAME data for each of the Cepheids over 
the thirteen epochs.}

\noindent{Table A10: -- $I$-band ALLFRAME data for each of the Cepheids over 
the nine epochs.}
\end{document}


\setcounter{page}{23}

\begin{deluxetable}{lcl}
\vspace{2in}
\tablecolumns{3}
\tablewidth{17cm}
\tablenum{4}
\tablecaption{Error Budget for NGC 4535}
\tablehead{
\colhead{Source of Uncertainty} & \colhead{Error (mag)} & \colhead{Comment}}
\startdata
{\bf CEPHEID PL CALIBRATION} & &  \nl
(a) LMC True Modulus & $\pm$0.10 & \nl
(b) V PL Zero Point & $\pm$0.05 & \nl
(c) I PL Zero Point & $\pm$0.03 & \nl \cline{2-2}
(S1) PL Calibration Uncertainty &$\pm$0.12&(a),(b),(c) combined in quadrature\nl
 & & \nl
{\bf HST CALIBRATION} & &  \nl
(d) HST V-Band Zero Point & $\pm$0.05 & \nl
(e) HST I-Band Zero Point & $\pm$0.05 & \nl\cline{2-2}
(S2) HST Calibration Uncertainty & $\pm$0.15 &(d),(e) coupled by reddening law \nl
 & & \nl
{\bf NGC$\,$4535 MODULUS} & & \nl
(f) PL Fit (V) & $\pm$0.05 & \nl
(g) PL Fit (I) & $\pm$0.04 & \nl\cline{2-2}
(R) Distance modulus Uncertainty & $\pm$0.05 &(f),(g) partially correlated \nl
 & & \nl
{\bf METALLICITY EFFECT} & & \nl
(M) Based on Kennicutt et al.~(1998,1999)& 0.17$\pm$0.11& \nl
 & & \nl
{\bf TOTAL UNCERTAINTY} & &  \nl
(R) Random Errors&$\pm$0.05 &\nl
(S) Systematic Errors&$\pm$0.26 &(S1),(S2),(M) combined in quadrature \nl
\enddata
\end{deluxetable}

\begin{deluxetable}{lcl}
\tablecolumns{3}
\tablenum{5}
\tablewidth{9cm}
\tablecaption{Distances to Virgo Cluster Galaxies}
\tablehead{
\colhead{Galaxy} & \colhead{Distance}  & \colhead{Reference} \\
\colhead{}     & \colhead{(Mpc)}       & \colhead{}}
\startdata
NGC$\,$4321  & $16.1\pm1.3$ & Ferrarese et al.~(1996) \nl
NGC$\,$4496A & $16.1\pm1.1$ & Saha et al.~(1996b)     \nl
NGC$\,$4535  & $16.0\pm1.9$ & This work               \nl
NGC$\,$4536  & $16.6\pm1.2$ & Saha et al.~(1996a)     \nl
NGC$\,$4548  & $16.0\pm2.0$ & Graham et al.~(1998)    \nl
NGC$\,$4571  & $14.9\pm1.2$ & Pierce et al.~(1994)    \nl
NGC$\,$4639  & $25.5\pm2.6$ & Saha et al.~(1997)      \nl
\enddata
\end{deluxetable}

\begin{deluxetable}{ccccc}
\tablecolumns{5}
\tablenum{A1}
\tablewidth{11cm}
\tablecaption{Coefficients used for ALLFRAME Photometry}
\tablehead{
\colhead{Chip} & \colhead{Filter}  & \colhead{Aperture}   & \colhead{WFPC2}
& \colhead{C}\\
\colhead{}     & \colhead{}        & \colhead{Correction} & \colhead{Zeropoint}
& \colhead{Coeff.}}
\startdata
 PC1 & V & $-0.093\pm0.086$ & 22.510 & $-1.036$\nl
 PC1 & I & $-0.102\pm0.067$ & 21.616 & $-1.940$\nl
 WF2 & V & $+0.021\pm0.023$ & 22.522 & $-0.971$\nl
 WF2 & I & $+0.028\pm0.016$ & 21.657 & $-1.796$\nl
 WF3 & V & $-0.003\pm0.042$ & 22.530 & $-0.956$\nl
 WF3 & I & $-0.026\pm0.043$ & 21.638 & $-1.809$\nl
 WF4 & V & $+0.030\pm0.031$ & 22.506 & $-0.984$\nl
 WF4 & I & $+0.005\pm0.043$ & 21.609 & $-1.895$\nl
\enddata
\tablecomments{The ALLFRAME instrumental magnitudes ($F555W$ and $F814W$ of
Eqns.~1-2) are obtained from the fitted magnitudes after applying the
correction $m_{\rm inst} = m_{\rm fit} + 2.5 * \log t + C$. The
coefficient $C$ equals $-25.00 + 0.016 + 2.5 * \log 4\ + AC + ZP$,
where $t$ is the exposure time, -25.00 is the ALLFRAME zeropoint,
+0.016 accounts for a pixel area normalization, $2.5 \log 4$ accounts
for the image conversion to integer data described in \S 2.1, $AC$ is
the aperture correction and $ZP$ is the WFPC2 zeropoint.}
\end{deluxetable}

\begin{deluxetable}{ccccr}
\tablecolumns{5}
\tablewidth{11cm}
\tablenum{A2}
\tablecaption{Comparison of DoPHOT and ALLFRAME photometry}
\tablehead{
\colhead{Set} & \colhead{Chip} & \colhead{$\Delta V\pm\sigma_V$} & \colhead{
$\Delta I\pm\sigma_I$} & \colhead{N}}
\startdata
Secondary & PC1 & $-0.003\pm0.047$ & $-0.052\pm0.074$ & 45 \nl
standards & WF2 & $+0.038\pm0.033$ & $+0.045\pm0.041$ & 45 \nl
          & WF3 & $+0.016\pm0.017$ & $+0.013\pm0.034$ & 45 \nl
          & WF4 & $-0.006\pm0.027$ & $+0.003\pm0.038$ & 45 \nl
          &     & & & \nl
\tableline
          &     & & & \nl
Cepheids & PC1 & $-0.015\pm0.029$ & $+0.030\pm0.019$ &  4 \nl
         & WF2 & $-0.080\pm0.065$ & $+0.033\pm0.150$ &  3 \nl
         & WF3 & $-0.037\pm0.087$ & $-0.047\pm0.087$ & 28 \nl
         & WF4 & $-0.029\pm0.093$ & $-0.051\pm0.093$ & 15 \nl
\enddata
\tablecomments{The magnitude differences are DoPHOT - ALLFRAME, and
the errors represent the {\it rms} scatter of the mean values.}
\end{deluxetable}%
